\documentclass[11pt]{article}
\usepackage[margin=1in]{geometry}
\usepackage{url}
\usepackage{setspace}
\usepackage{subcaption} 
\usepackage{graphicx}
\usepackage{caption}
\usepackage{amsmath}
\usepackage{algpseudocode} 
\usepackage{algorithm}
\usepackage{authblk}

\usepackage{soul} 
\usepackage{xcolor} 

\soulregister\cite7
\soulregister\ref7
\soulregister\etal7

\newcommand{\etal}{et\unskip~al.\@}
\newcommand{\ie}{i.e.~\@}

\begin{document}
\title{Learning a model of shape selectivity in V4 cells reveals shape encoding mechanisms in the brain}
\author{Paria Mehrani and John K. Tsotsos}
\affil{Department of Electrical Engineering and Computer Science, \\
York University, Toronto, Canada}
\affil{\{paria, tsotsos\}@eecs.yorku.ca}
\date{}
\maketitle
\begin{abstract}
	The mechanisms involved in transforming early visual signals to curvature representations in V4	are unknown. We propose a hierarchical model that reveals V1/V2 encodings that are essential components for this transformation to the reported curvature representations in V4. Then, by relaxing the often-imposed prior of a single Gaussian, V4 shape selectivity is learned in the last layer of the hierarchy from Macaque V4 responses. We found that V4 cells integrate multiple shape parts from the full spatial extent of their receptive fields with similar excitatory and inhibitory contributions. Our results uncover new details in existing data about shape selectivity in V4 neurons that with further experiments can enhance our understanding of processing in this area. Accordingly, we propose designs for a stimulus set that allow removing shape parts without disturbing the curvature signal to isolate part contributions to V4 responses.

\end{abstract}
\allowdisplaybreaks


\section{Introduction}
Transformation of the shape signal in the ventral stream from low-level visual representations of V1 (oriented edges)~\cite{hubel1959receptive,hubel1963visual} and V2 (corners and junctions)~\cite{V2_ito2004representation,V2_hegde2000} to more abstract representations in IT (objects, faces, etc.)~\cite{tanaka1996IT, kobatake1998IT} is still unknown. V4, as an intermediate processing stage in this pathway and as the major source of input to IT, is believed to play a role in this transformation~\cite{Roe2012V4,pasupathy2020visual}. Yet, selectivities to shape features in V4 neurons add to this mystery. Specifically, with many V4 neurons selective to convex and concave shape parts at a specific position relative to the object center~\cite{Pasupathy99_contour_features,pasupathy2001_shape_representation,pasupathy2002_nature}, it remains unclear how such part-based and object-centered curvature representations are achieved in the ventral stream. Here, we propose a mechanistic hierarchical network that explicitly models neurons in V1, V2 and V4 and could explain this transformation. In particular, we break this problem into two sub-problems tackled separately: first, transformation of the shape signal into an object-center curvature encoding in the ventral stream; second, part-based selectivity in V4.

\begin{figure}[tp!]
	\centering
	\begin{subfigure}{0.57\textwidth}
		\centering
		\includegraphics[width=\textwidth]{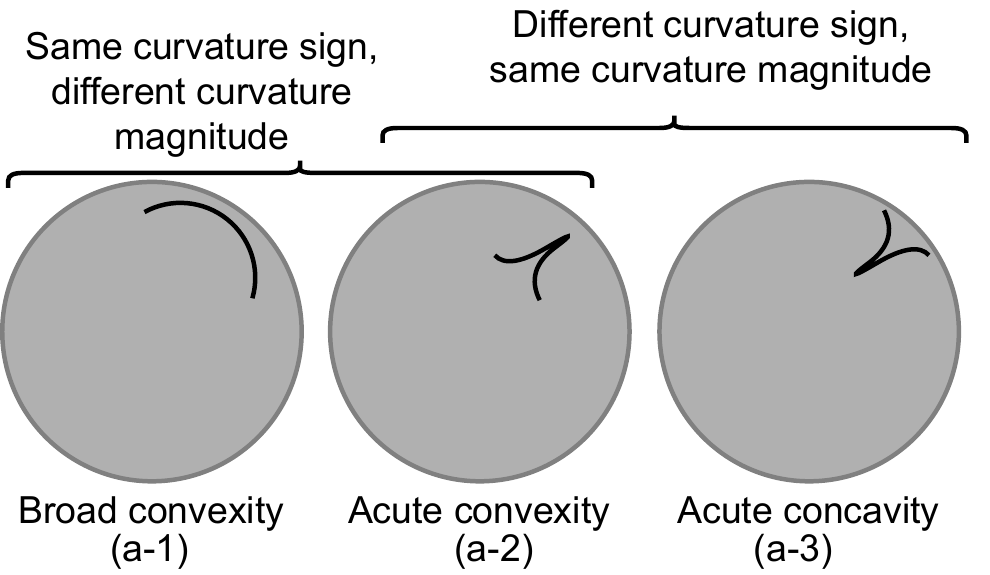}
		\caption{}
		\label{subfig:Pasupathy_fig5}
	\end{subfigure}~\hspace{10pt}
	\begin{subfigure}{0.3\textwidth}
		\centering
		\includegraphics[width=\textwidth]{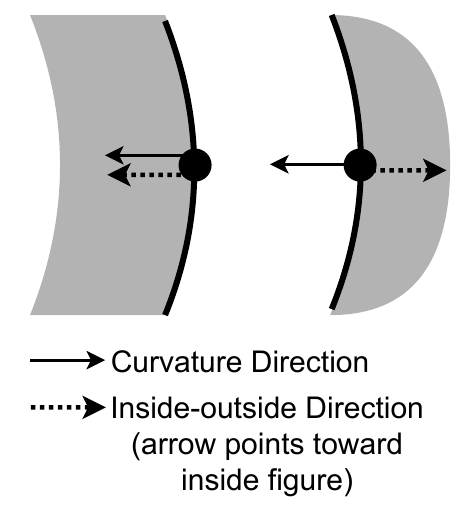}
		\caption{}
		\label{subfig:curv_sign_two_arrows}
	\end{subfigure}\\ \vspace{0.5cm}
    \begin{subfigure}{0.6\textwidth}
    	\centering
    	\includegraphics[width=\textwidth]{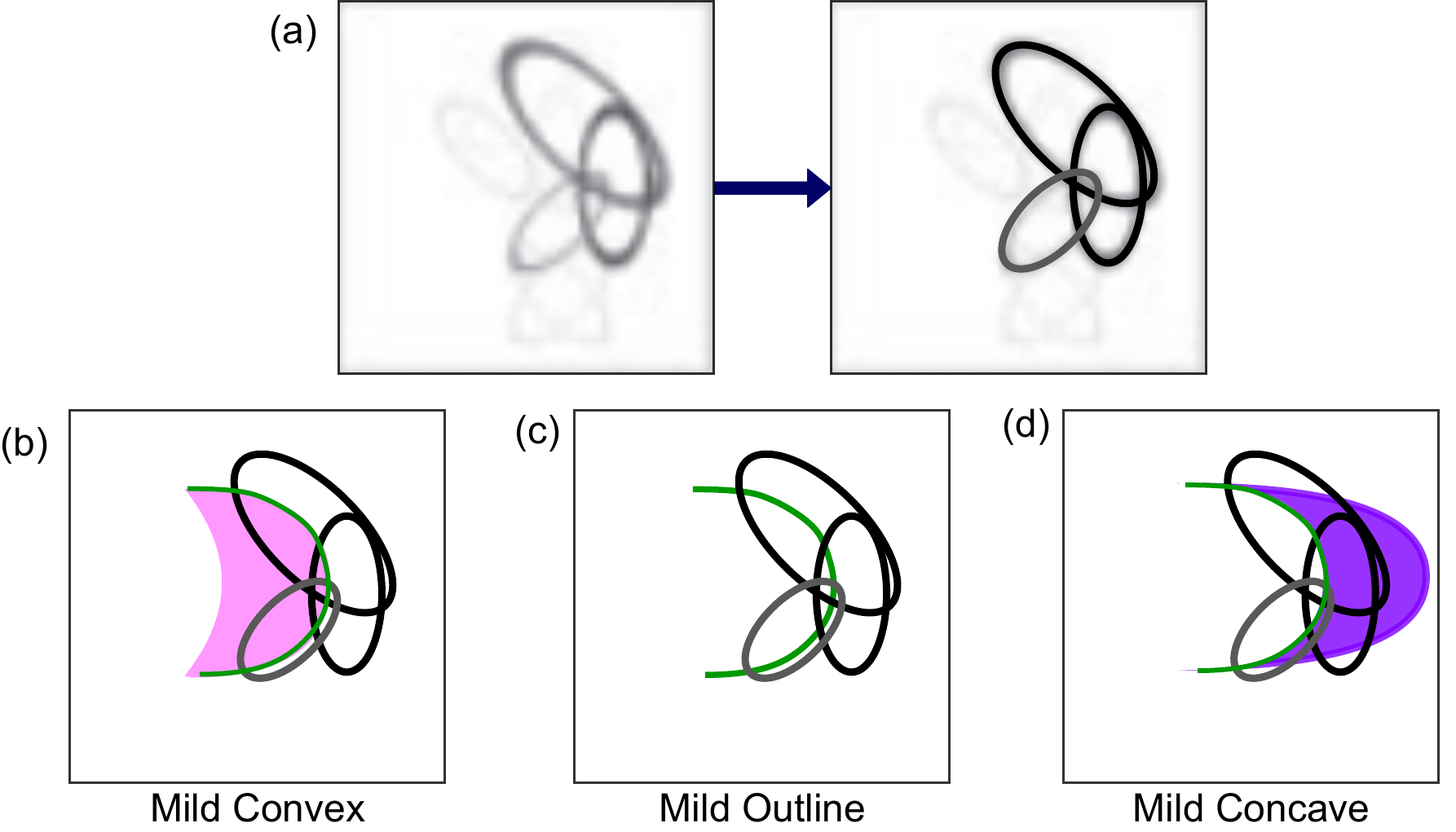}
    	\caption{}
    	\label{fig:HMAX_FG_example}
    \end{subfigure}
	\caption{Signed curvature selectivity in V4. (\subref{subfig:Pasupathy_fig5}) Signed curvature encapsulates two curvature components: magnitude and sign. Here, curved segments in (a-1) and (a-3) share either curvature magnitude or sign with the one in (a-2). A V4 cell selective to acute convexities on the upper right side of its RF will be strongly activated by the stimuli in (a-2) and not (a-1) or (a-3). (\subref{subfig:curv_sign_two_arrows}) Conventionally, inside-outside direction for a closed simple planar shape defines a unique signed curvature for each point on its bounding contour. Even with identical curvature magnitude and direction, the same curved segment might be a convex or a concave part of a closed shape. One such example is demonstrated here for the two gray shapes with identical curve segments that are denoted with solid black curves. For a simple closed shape, with matching curvature direction and inside-outside direction, the curve is considered convex. Otherwise, it is concave. 
	(\subref{fig:HMAX_FG_example}) Lack of curvature sign representations in previous V4 models. 
		A recovered V4 RF by HMAX is shown in the left square on the top row (extracted from~\cite{cadieu2007model}). Each oval represents an orientation-selective C1 afferent from the HMAX hierarchy (see~\cite{cadieu2007model} for details) with its perimeter intensity indicating its weight to the simulated cell responses. 
	For simplicity and to demonstrate a lack of curvature sign representation in this neuron, we retained three components with largest weights, \ie, strongest contributions as shown in the right square. Panels in the bottom row show the RF profile super-imposed with stimuli similar to those employed in~\cite{Pasupathy99_contour_features}. Specifically, the green curve is identical in all three stimuli but as a mild convex, mild outline and mild concave segment from left to right. V4 neurons distinguish between these three stimuli~\cite{Pasupathy99_contour_features}. This HMAX neuron, however, will exhibit strong activations to all three stimuli as the green curved segment is aligned and overlaps with the three orientation-selective C1 afferents resulting in strong responses.}
	\label{fig:signed_curvature}
\end{figure}

\textbf{\underline{Problem 1:} Signed curvature encoding.} Findings of shape processing in V4 provide evidence for selectivity to two curvature components: curvature magnitude (deviation from a straight line) and curvature sign (convexity-concavity, determined according to a set origin), together defining the scalar called \textit{signed curvature}. Figure~\ref{fig:signed_curvature}(\subref{subfig:Pasupathy_fig5}) depicts how two curve segments can share the same curvature sign with different curvature magnitudes or have the same curvature magnitude but different curvature signs. A V4 neuron selective to acute convexities at the top right side of its receptive field (RF) exhibits strong responses to stimuli in Figure~\ref{fig:signed_curvature}(a-2) and not to those on either side (See~\cite{pasupathy2001_shape_representation}, Figure 5 for such a tuning in a Macaque V4 cell). 

The angular position and curvature (APC) model introduced by Pasupathy~\etal~\cite{pasupathy2001_shape_representation} revealed signed curvature selectivity in V4 neurons. Despite the observed selectivities to both curvature magnitude and sign, existing models of V4~\cite{cadieu2007model,Antonio_2012,hu2014sparsity,wang2016modeling,eguchi2015computational,wei2015v4RBM,antonio2016RBM,hatori2016sparse,kawakami2020cell} encode curvature magnitude but none models curvature sign. Lack of a curvature sign encoding in these models result in responses that are in contrast to the reported observations in V4. Figure~\ref{fig:signed_curvature}(\subref{fig:HMAX_FG_example}) gives a simple example to demonstrate this disparity. Additionally, when the goal is to understand the development of a signed curvature representation in the ventral stream, these models leave a gap in our understanding of shape signal transformation from orientation encodings to selectivity to a signed curvature representation. This gap is especially evident in the APC model with direct mapping of stimulus shape to the position and signed curvature domain. To address this problem, we introduce a hierarchical network, dubbed SparseShape, that explicitly models both curvature magnitude and sign. Hence, a signed curvature encoding is achieved. We modeled the curvature sign representation according to the observation that it can be uniquely determined for any point on the bounding contour of a simple planar closed shape by combining two signals: inside-outside and curvature direction, \ie, the direction toward which the contour curves (Figure~\ref{fig:signed_curvature}(\subref{subfig:curv_sign_two_arrows})). Neural correlates for both signals in the ventral stream, namely, border ownership (BO)~\cite{BOwn_Zhou} and endstopping~\cite{V2_dobbins1987endstopped,V2_ito2004representation}, support the plausibility of the proposed model. 

\begin{figure}[tp!]
	\centering
	\begin{subfigure}{0.9\textwidth}
		\centering
		\includegraphics[width=\textwidth]{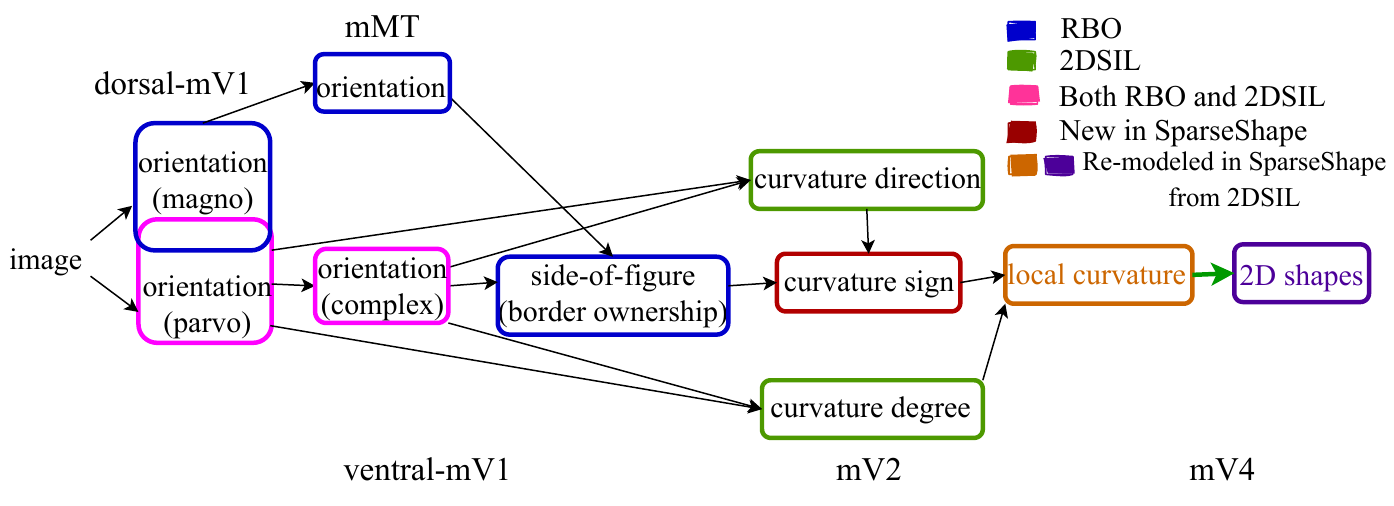}
		\caption{}
		\label{subfig:hierarchy}
	\end{subfigure}\\ \vspace{0.5cm}
    \begin{subfigure}{0.78\textwidth}
    	\centering
    	\includegraphics[width=\textwidth]{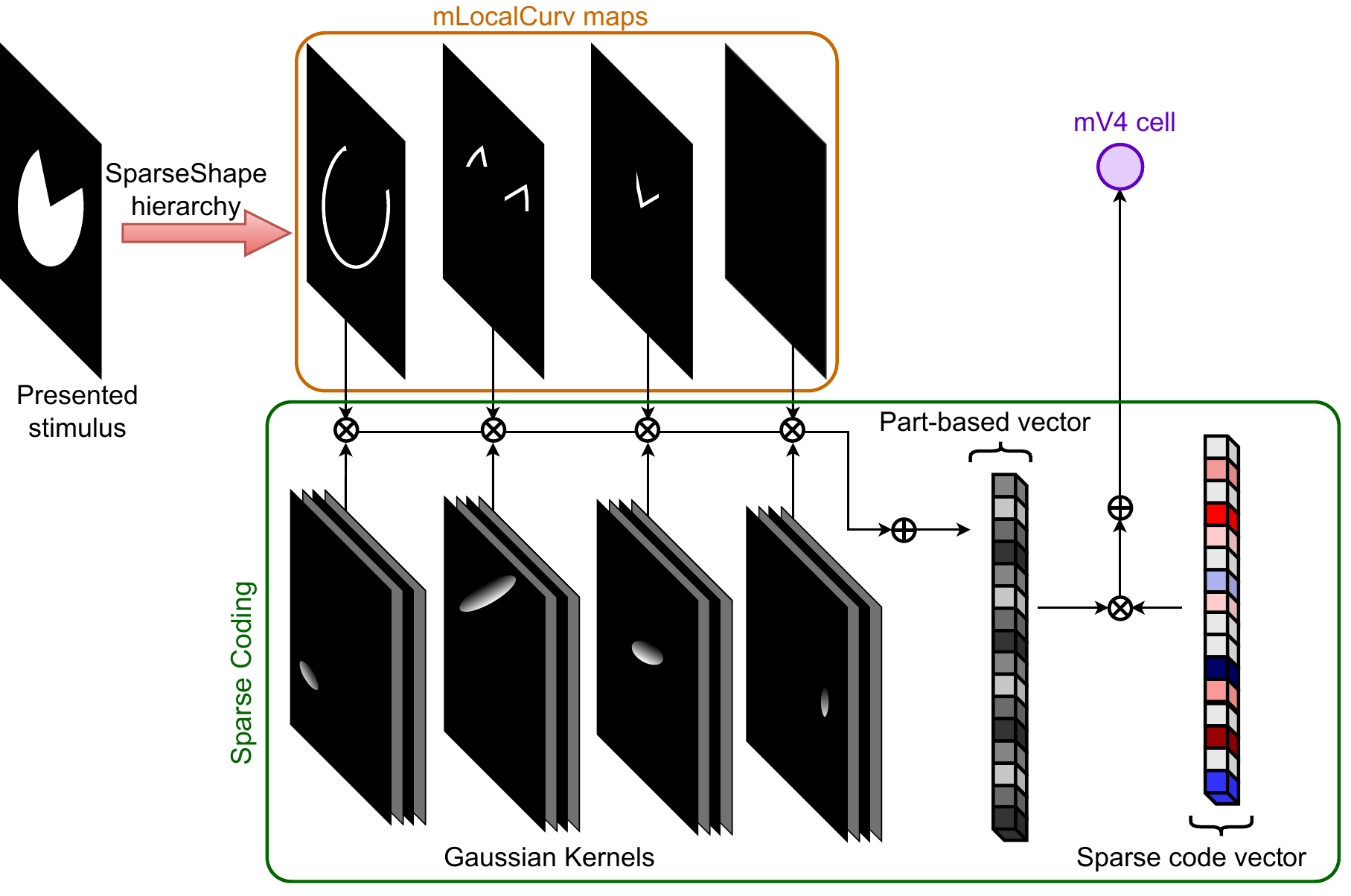}
    	\caption{}
    	\label{subfig:sparse_model}
    \end{subfigure}
	\caption{The proposed model. (\subref{subfig:hierarchy}) The SparseShape hierarchy combines and extends two previously-introduced hierarchical models, RBO~\cite{RBO} and 2DSIL~\cite{Antonio_2012}. The color of each box represents the origin of the neuron from 2DSIL, RBO, or SparseShape. 2DSIL models curvature magnitude and direction through endstopped neurons. To achieve a signed curvature encoding, inside-outside information is represented by incorporating the RBO network modeling border ownership. In SparseShape, new neurons modeling curvature sign are added to the network that enable signed curvature selectivity in the mV4 layer. Accordingly, the model local curvature cells are re-modeled with feedforward signals from curvature degree and sign cells. Similarly, mV4 cells are re-modeled with the sparse coding step. (\subref{subfig:sparse_model}) The various components of the supervised sparse coding algorithm are shown in the green box in this figure corresponding to the green arrow in panel (\subref{subfig:hierarchy}). Given mLocalCurv maps and Macaque V4 responses to stimuli in the training set, the algorithm looks for a combination of shape conformations that could best explain mV4 responses. To account for the various positions of shape parts, a set of 9 Gaussian kernels spanning a $3\times3$ grid over the RF filter mLocalCurv responses at each cell. The filtered responses form the part-based vector. The sparse code vector denotes the weight of each part contributing to the responses. The parameters of the Gaussian kernels are fixed while the sparse code vector is learned in this step.}
	\label{fig:model_architecture}
\end{figure}

SparseShape, whose architecture is depicted in Figure~\ref{fig:model_architecture}(\subref{subfig:hierarchy}), combines and extends two previous models~\cite{RBO,Antonio_2012}. In SparseShape, simple and complex cells extract oriented edges. Complex cell responses modulated by early recurrence from the dorsal stream result in border ownership (See~\cite{RBO} for details). Combining simple and complex cell responses result in two types of endstopped neurons: curvature degree and curvature direction (details explained in~\cite{Antonio_2012}). Curvature degree cells at four scales represent curvature magnitude. Curvature sign encoding is achieved by combining BO responses with that of curvature direction neurons. Combining curvature sign and curvature degree signals yield the signed curvature representation manifested in model local curvature neurons. In addition to support from available biological findings, the three types of endstopped curvature cells (degree, direction and sign) that give rise to a signed curvature encoding have geometric interpretations that are explained in Section~\ref{section:Methods}. Finally, model local curvature cells feed their signal to model shape-selective cells in the output layer of SparseShape representing V4 selectivities. The SparseShape network meets known biological properties with its parameters set according to neurophysiological findings.

\textbf{\underline{Problem 2:} Part-based selectivity.} Pasupathy~\etal~\cite{pasupathy2001_shape_representation} fit boundary conformation tunings in V4 cells with a Gaussian function in the APC space.
Nonetheless, because they imposed the strong prior of fitting a single Gaussian to V4 responses, complex and long-range interactions between shape parts within the RF could not be captured. Long-range interactions were accounted for in HMAX~\cite{cadieu2007model} and 2DSIL~\cite{Antonio_2012} by heuristic algorithms that determined shape components with optimizing the model fit to Macaque V4 responses. All three approaches, however, were restricted to facilitatory contributions of shape parts; another prior imposed on the recovered tunings. Whereas neurons in other brain areas exhibit a combination of excitatory and inhibitory regions within their RFs, these models chose to only represent excitation. We investigated the potential impact of including inhibition. 
Specifically, given local curvature cell responses in SparseShape to shape stimuli, for each Macaque V4 neuron, we utilized a supervised sparse model and learned a set of contour segments that determine the neuron responses (Figure~\ref{fig:model_architecture}(\subref{subfig:sparse_model})). This step is discussed in detail in Section~\ref{section:Methods}. 

The learned representations suggest that V4 cells combine shape parts from the full extent of their RF with both facilitatory and inhibitory contributions. Probing our model cell responses to variations in scale and position, using data with which the model was not trained, we found our model V4 neurons replicate the invariance reported in biological V4 cells~\cite{elShamayleh2016scale}. These findings provide additional details that were uncovered in existing V4 data, and together with the proposed hierarchy provide new insights into shape processing in the ventral stream. Moreover, our findings suggest new experimental paradigms that could further enhance our understanding of V4 selectivities.
 
 In what follows, we will make a distinction between our model and brain areas by referring to those as layers and areas respectively. That is, a set of model neurons implementing cells in a brain area will be referred to as a model layer. For example, model layer V2 implements cells in brain area V2. Moreover, whenever a brain area name is preceded with ``m'', it is referring to the corresponding model layer, for example, mV2 for model layer V2. 
 
 \section{Results}
 We conducted experiments with two sets of stimuli: 366 parametric shapes combining convex-concave parts into closed shapes borrowed from~\cite{pasupathy2001_shape_representation} and parametric shapes with a single varying part along its contour designed according to the set from~\cite{elShamayleh2016scale}. The latter set was employed to test invariance in mV4 responses. The former shape set, which we call the standard set, was utilized in all other experiments as well as training the model. The standard set shapes were scaled such that stimulus edges were offset from the RF center by $0.75\times$ RF size. The shapes in the invariance experiments were scaled following~\cite{elShamayleh2016scale}. The SparseShape network input was set to $400\times400$ pixels and we measured $1^\circ$ visual angle at 50cm to be $32\times32$ pixels. We set mV4 RFs at $4^\circ$ following~\cite{felleman1991distributed} equivalent to $128\times128$ pixels.
 
 Supervised learning in the last layer of SparseShape was performed with electrophysiological responses of 109 Macaque V4 neurons reported in~\cite{pasupathy2001_shape_representation} provided to us by Dr. Pasupathy. That is, for each Macaque V4 cell, we assigned its responses to a mV4 neuron and learned the model weights from mLocalCurv cells to the mV4 neuron. A stratified division of the standard shape set into $60\%-40\%$ train and test splits was performed for each individual neuron to ensure an assortment of responses were present in each set. The hyper-parameters for learning the last layer weights were determined using cross-validation over the training set followed by learning the weights with all the training shapes. All other model parameters were set according to experimental findings following~\cite{Antonio_2012, RBO}.
 
 \subsection{Model comparison}
 To measure model performance, we report Pearson's $r$ correlation coefficients and mean absolute error (MAE) between model and Macaque V4 responses, separately for train and test sets. Additionally, we compare our model performance against two previous models for V4, namely, 2DSIL~\cite{Antonio_2012} and HMAX~\cite{cadieu2007model}. These two models are closely related to our proposed model with direct comparison with Macaque V4 data. Both 2DSIL and HMAX modeled curvature magnitude and direction and not curvature sign. Both models utilized heuristic approaches to determine weights representative of V4 selectivities in their hierarchies. HMAX employed a greedy algorithm to recover V4 RFs (See~\cite{cadieu2007model}, Figure 9) while shape templates in 2DSIL were computed based on common shape parts among each neuron's preferred stimuli. In HMAX, the recovered RFs do not necessarily result in curvature-like configurations whereas in 2DSIL, having a shape template is not guaranteed (they reported results for 75 out of 109 Macaque V4 cells). 
 
 Alongside HMAX and 2DSIL, we compare our model performance with that of the APC model~\cite{pasupathy2001_shape_representation}. 
 APC represents signed curvature and thus is most similar to SparseShape. In addition to fitting the tunings in the two-dimensional APC space, Pasupathy~\etal~\cite{pasupathy2001_shape_representation} evaluated a four-dimensional APC model (3 curvature and 1 angular-position) by considering the two neighboring curvature components on either side of a shape part. We call these two versions as APC--2D and APC--4D in what follows. Finally, we compare SparseShape with the correlations of 100 random neurons selected from convolutional layer 2 and fully-connected layer 7 of Alexnet~\cite{alexnet} reported by Pospisil~\etal~\cite{pospisil2018artiphysiology}.
 
 \subsection{Shape part selectivity}
\begin{figure}[!t]
	\centering
	\begin{subfigure}{\textwidth}
		\centering
		\includegraphics[width=\textwidth]{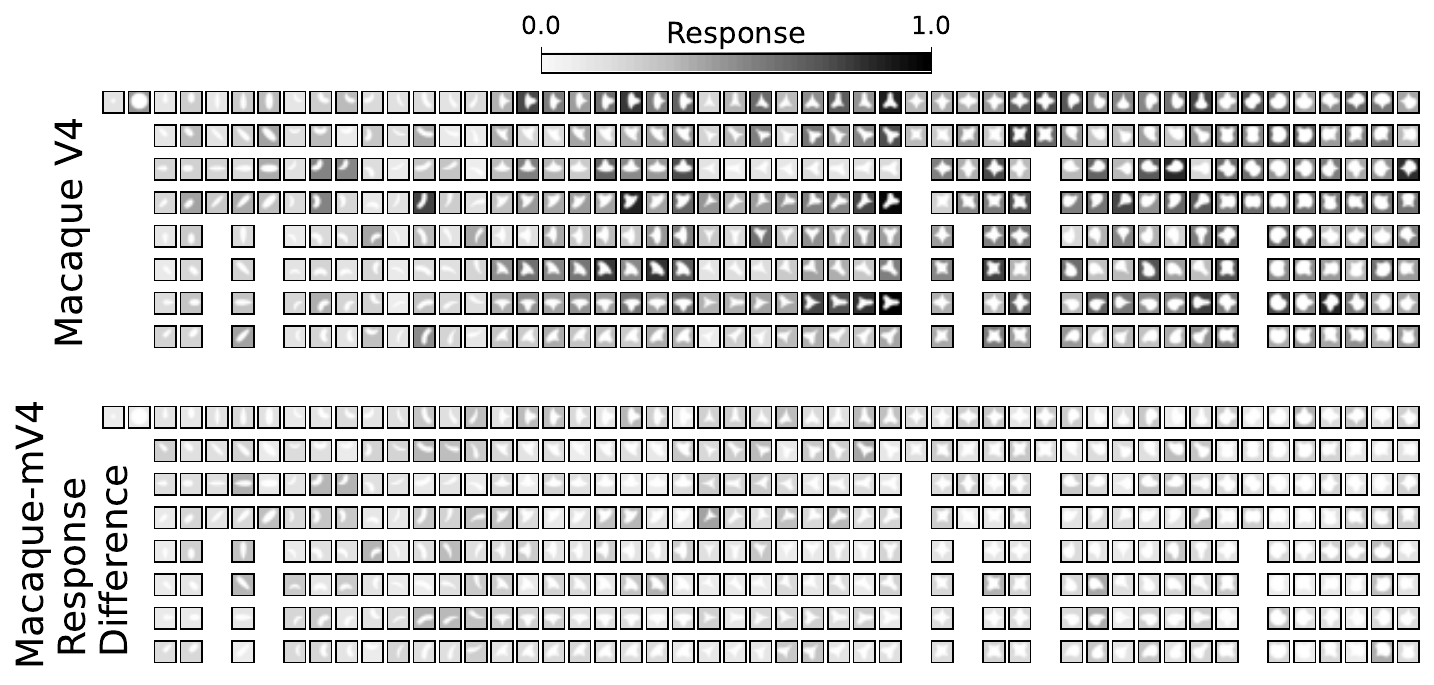}
		\caption{}
		\label{subfig:V1601_resp}
	\end{subfigure}
	\begin{subfigure}{0.42\textwidth}
		\centering
		\includegraphics[width=\textwidth]{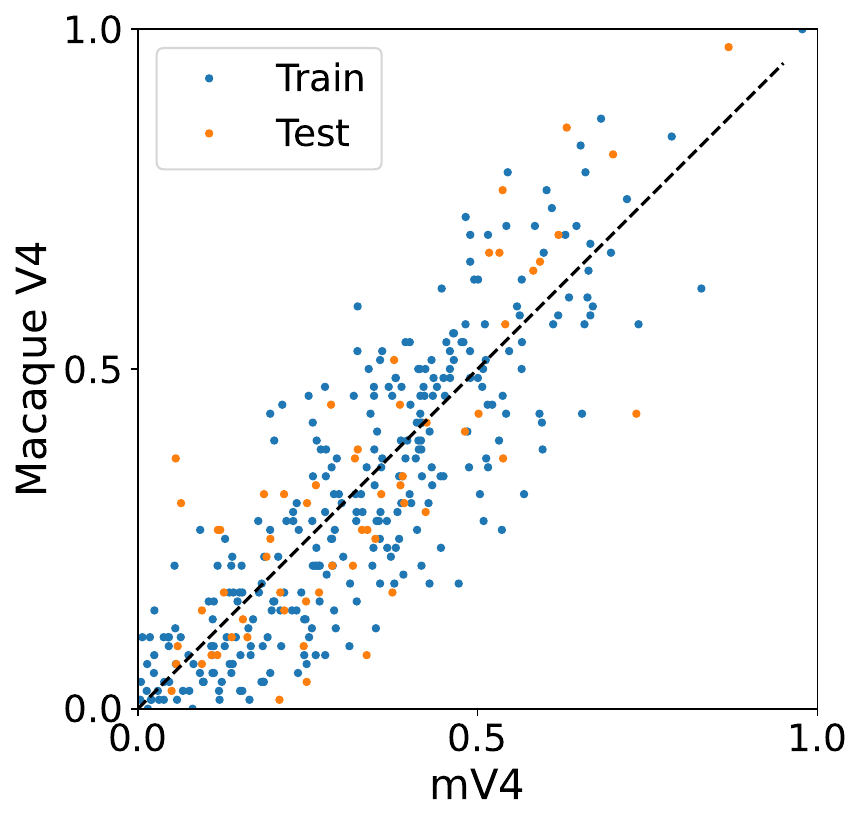}
		\caption{}
		\label{subfig:V1601_corr}
	\end{subfigure}
	\caption{Responses of a learned mV4 cell.~(\subref{subfig:V1601_resp}) Two rows Macaque V4 responses and response differences with mV4 to the stimulus set introduced by Pasupathy~\etal~\cite{pasupathy2001_shape_representation}. Each shape is depicted as a white icon in a square. The intensity of the ground in each square indicates the response strength. Small response differences in the second row demonstrate a similar pattern of responses in the mV4 cell and the Macaque V4 cell.~(\subref{subfig:V1601_corr}) Macaque V4 responses plotted against mV4 activations illustrates a strong correlation between these responses ($r=0.86$ and $r=0.83$ for train and test shapes respectively).}
	\label{fig:V1601_resp_corr}
\end{figure}

\begin{figure}[!hp]
	\centering
	\begin{subfigure}{\textwidth}
		\centering
		\includegraphics[width=\textwidth]{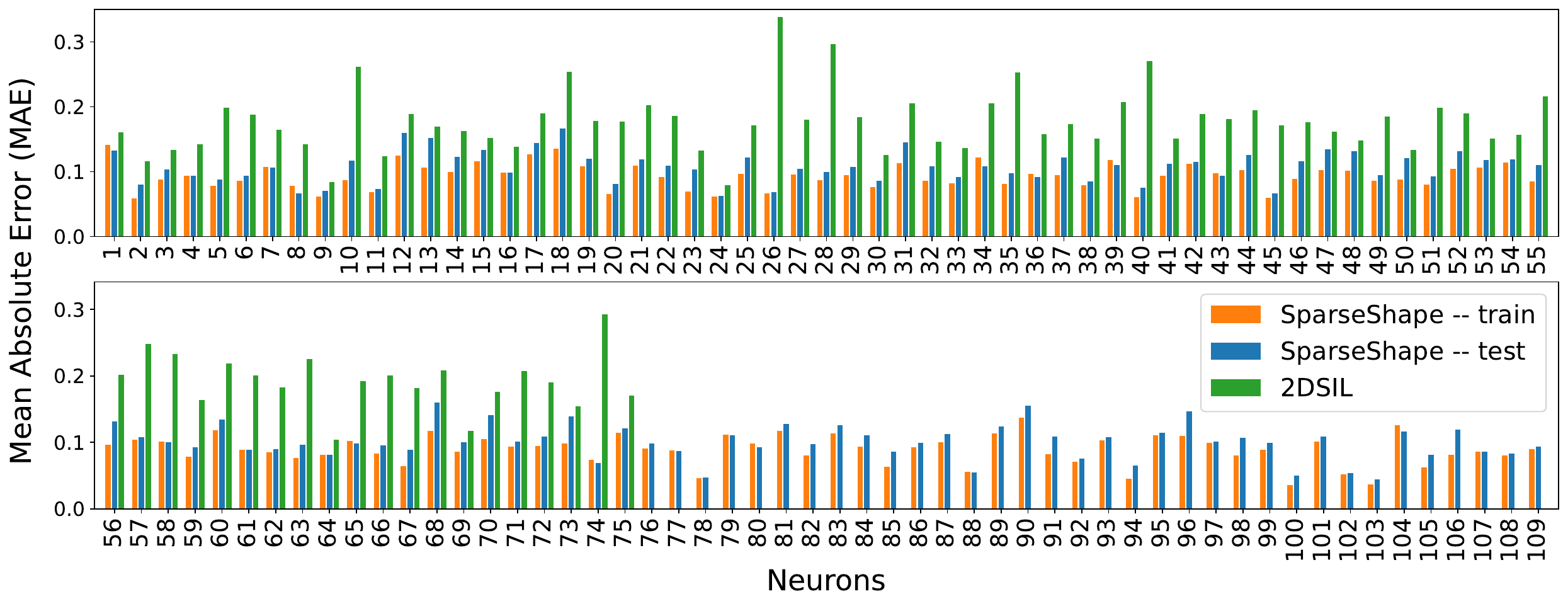}
		\caption{}
		\label{fig:mean_error_bars}
	\end{subfigure}\\
    \begin{subfigure}{\textwidth}
    	\centering
    	\includegraphics[width=\textwidth]{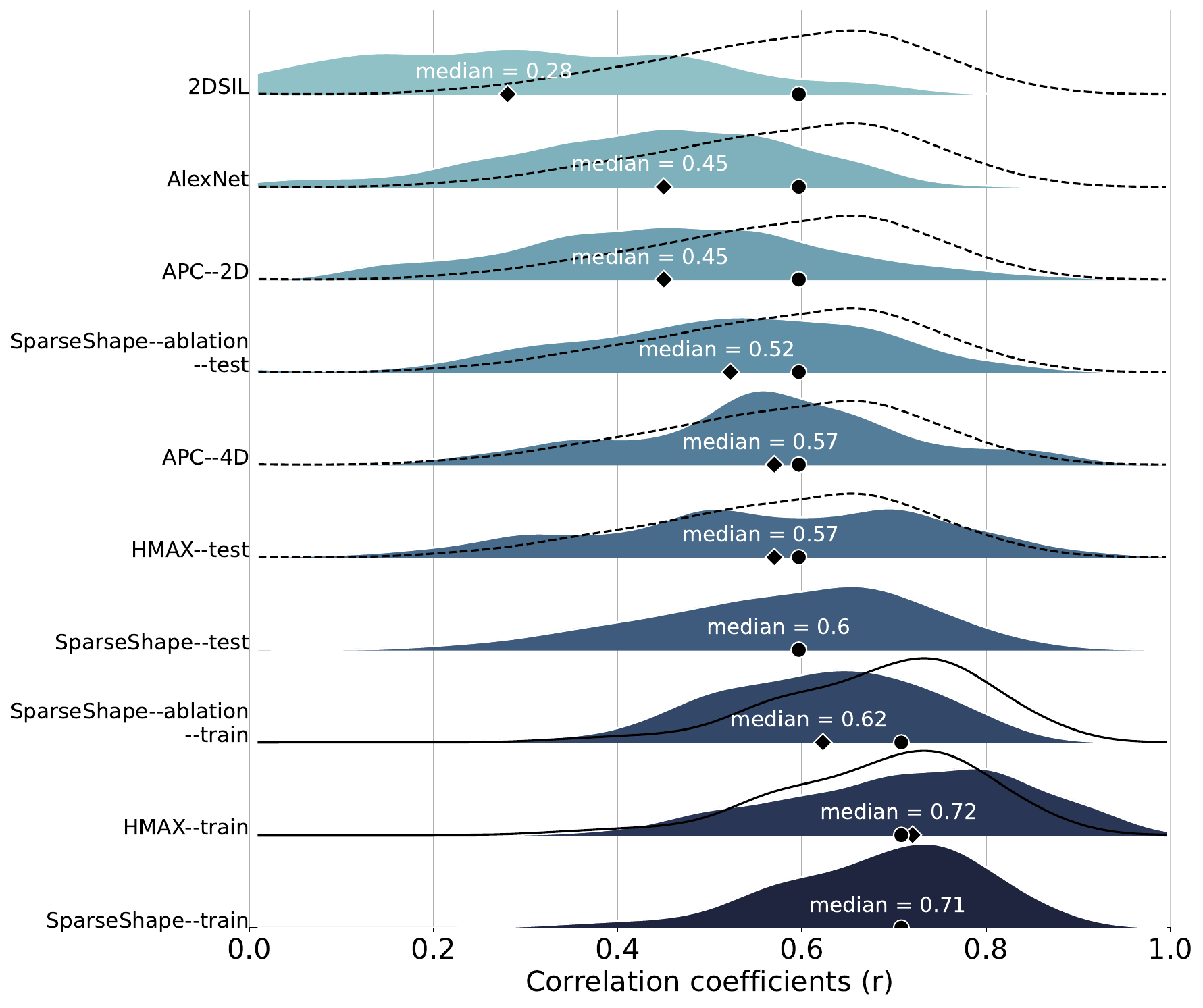}
    	\caption{}
    	\label{fig:correlation_comparison}
    \end{subfigure}
    \caption{Caption on the next page}
    \label{fig:error_correlation_comparison}	
\end{figure}

Figure~\ref{fig:V1601_resp_corr}(\subref{subfig:V1601_resp}) depicts responses of a Macaque V4 neuron to the standard shape set and response differences between Macaque V4 and its corresponding mV4 cell. Small response differences suggest similar selectivities in Macaque V4 and mV4 cells that are also confirmed with the strong correlations demonstrated in Figure~\ref{fig:V1601_resp_corr}(\subref{subfig:V1601_corr}). 
 To measure this similarity in the population, we computed the MAE separately for train and test shapes for all 109 Macaque V4 cells, illustrated in Figure~\ref{fig:error_correlation_comparison}(\subref{fig:mean_error_bars}). This figure demonstrates a decrease of MAE in SparseShape versus 2DSIL in 75 Macaque V4 neurons. Additionally, the remaining 34 cells follow a similar trend in train and test MAE. The average MAE over all mV4 cells is 0.09 in SparseShape versus 0.18 in 2DSIL. MAEs for the rest of the models were not available and therefore, these models are missing in Figure~\ref{fig:error_correlation_comparison}(\subref{fig:mean_error_bars}).
\addtocounter{figure}{-1}
\begin{figure}[!t]
	\caption{Model performance evaluation. (\subref{fig:mean_error_bars}) Mean absolute error in responses of model V4 cells for 2DSIL and SparseShape. Neuron errors are plotted in two rows for better visualization. SparseShape improves the performance measured as the mean difference of responses with biological V4 cells compared to 2DSIL. (\subref{fig:correlation_comparison}) Population distribution of correlation coefficients, $r$, for various V4 models plotted as a ridgeline plot in separate rows. Each model is indicated next to its distribution. The models are sorted according to their medians from bottom to top, with the exception of SparseShape--train. For easier comparison, SparseShape--train distribution is plotted as a solid line for HMAX--train and SparseShape--ablation--train rows and SparseShape--test distribution as a dotted curve in rows for all other models. The median for each distribution is denoted by a diamond shape, the value of which is written in white, and the median of Sparseshape--train/test distributions are identified with a black circle in each row.  This plot demonstrates that SparseShape has comparable distributions to HMAX with better generalization capacity over the standard shape set and improves the correlation coefficients compared to other models. This plot also includes correlation distributions of an ablation study where we limited the contribution type in SparseShape to facilitatory ones to examine the effects of inhibitory shape parts to V4 responses. Comparing SparseShape and SparseShape--ablation distributions suggest inhibitory shape parts integration in V4. Among all the models, Alexnet and 2DSIL with smaller correlations compared to the rest of the models demonstrate poor fit of their neurons to V4 data.} 
\end{figure}

We computed the correlation coefficients, $r$, as a goodness-of-fit measure between Macaque and model cell responses following~\cite{pasupathy2001_shape_representation, cadieu2007model}. While the APC model was fit based on responses to all the 366 stimuli, HMAX employed a six-fold cross-validation and reported the average correlation coefficient. In other words, HMAX fits six models to each V4 cell and reports the average performance of the six models. Cross-validation correlations are not appropriate indicators of the generalization abilities of the model and with six models learned for each neuron, it is unclear which must be considered as the true model of the cell. However, in absence of generalization data from HMAX, we compare SparseShape with their reported cross-validation correlations. Note that SparseShape evaluation is based on separate train-test splits that were explained earlier. 
Figure~\ref{fig:error_correlation_comparison}(\subref{fig:correlation_comparison}) illustrates the correlation coefficient distributions for APC--2D, APC--4D (extracted from~\cite{pasupathy2001_shape_representation}, Figure 9), HMAX (extracted from~\cite{cadieu2007model}, Figure 5), Alexnet (extracted from~\cite{pospisil2018artiphysiology}, Figure 14), 2DSIL and SparseShape. A ridgeline plot separates theses distributions. For easier comparison, SparseShape--train/test distributions are plotted as solid and dotted curves respectively in rows of other models.

Despite the slight shift in the correlations distribution for HMAX--train compared to SparseShape--train depicted in Figure~\ref{fig:error_correlation_comparison}(\subref{fig:correlation_comparison}), both models have comparable medians at 0.72 and 0.71, indicating a great overlap between the simulated populations. In contrast, SparseShape--test distribution is slightly shifted with a larger median at 0.6 compared to 0.57 for HMAX--test, demonstrating better generalization ability over the standard set for SparseShape. The impressive correlations and recovery of V4 selectivities that HMAX displays are based on the assumption that there are no intermediate computational stages between orientation (complex) neurons and 2D shape cells. This is in line with the way computer vision has viewed the problem of 2D shape over the past; it is not consistent with the available neurobiology, however, that shows a much richer connectivity (See~\cite{ungerleider2008V4}) that would suggest this assumption is over-simplifying the network. This has a number of consequences. As reported in~\cite{bair2015modeling}, HMAX neurons are not position-invariant whereas biological V4 cells do exhibit shape position invariance. An intermediate step of curvature sign would have provided what is required to make these cells position invariant, as our model shows. Second, the assumption leads to HMAX being a single-task network: it may provide a good fit to a particular set of V4 data, but when situated in a full network (such as detailed by~\cite{ungerleider2008V4}, and many others), HMAX alone may not play the full role for which those neurons are responsible. Other perceptual functions such as border ownership, figure-ground segmentation, localizing concave or convex shape elements, recognizing sharpness of shape, and more would fall outside the range of HMAX.

Interestingly, Alexnet with a deep convolutional neural network architecture is outperformed by all other models except for 2DSIL and APC--2D, demonstrating poor fit of its units to V4 data.

The APC distributions are most interesting. Despite the fit to the whole standard set, both APC--4D and APC--2D distributions are shifted to smaller correlations compared to the SparseShape. Larger correlation coefficients in APC--4D versus APC--2D, as pointed out in~\cite{pasupathy2001_shape_representation}, suggest complex configuration encodings in V4 cells. Likewise, comparing APC--4D and SparseShape distributions suggests a complicated pattern of interactions between shape parts within V4 RFs, beyond a combination of three adjacent parts.


 \subsection{Sparsity}
\begin{figure}[!h]
 	\centering
 	\begin{subfigure}{0.43\textwidth}
 		\centering
 		\includegraphics[width=\textwidth]{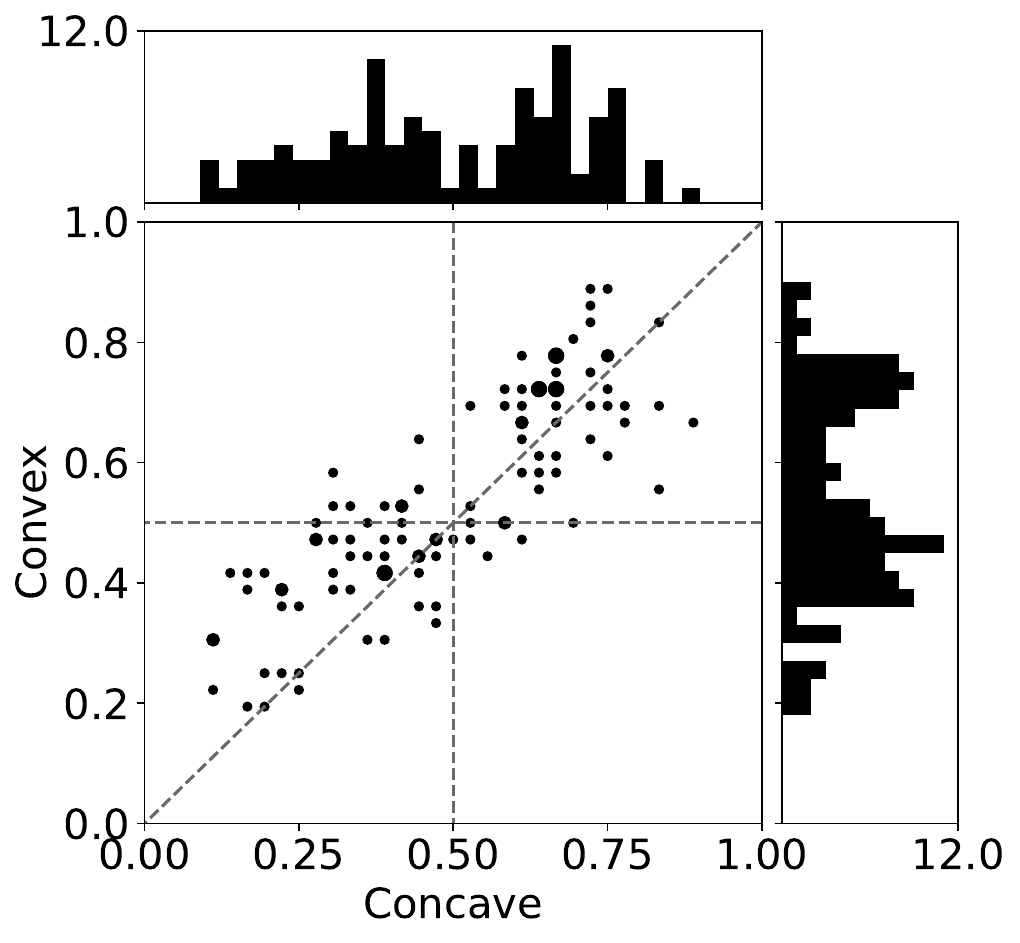}
 		\caption{}
 		\label{fig:sparse_percentage}
 	\end{subfigure}~
    \begin{subfigure}{0.56\textwidth}
    	\centering
    	\includegraphics[width=\textwidth]{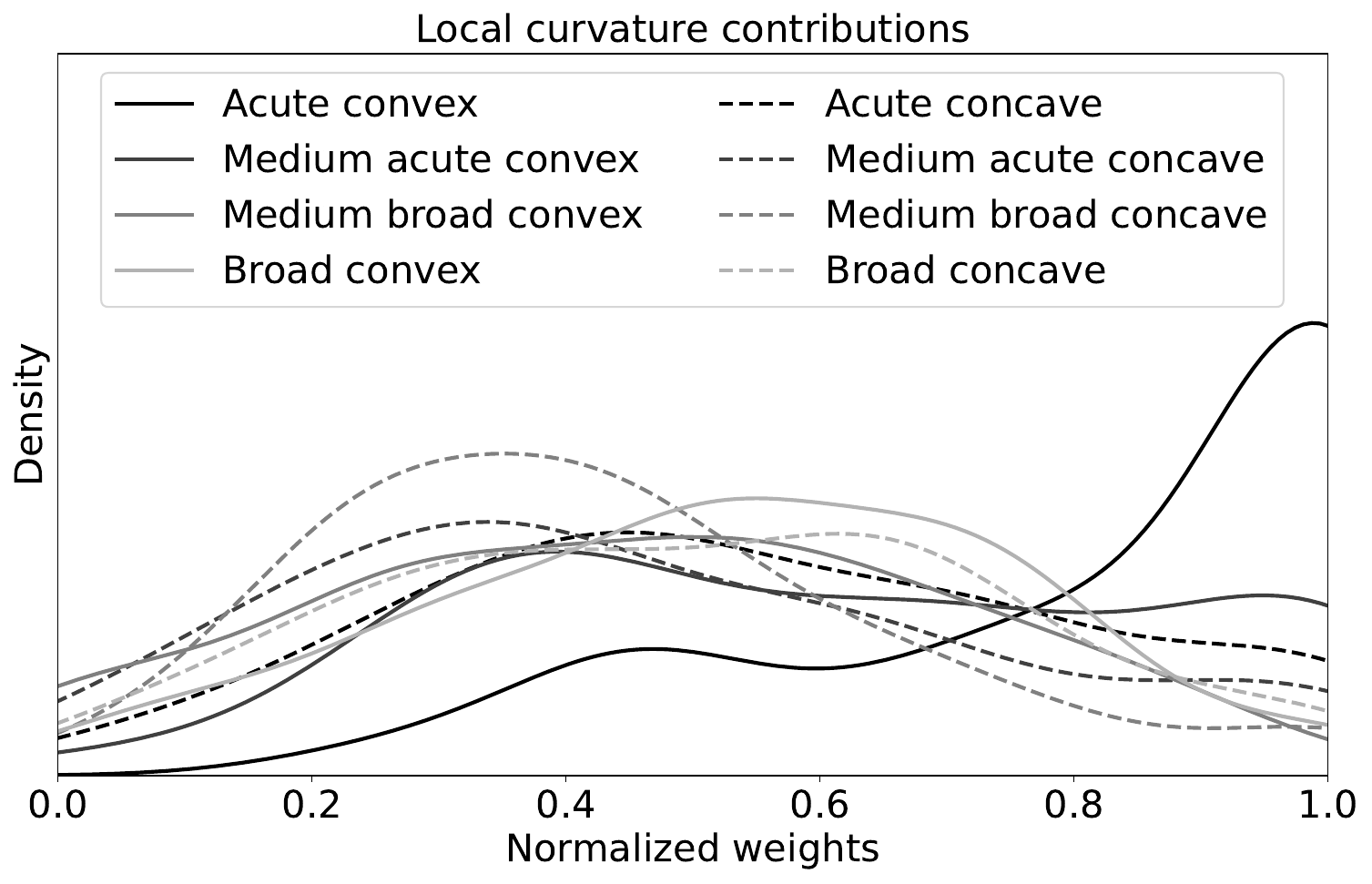}
    	\caption{}
    	\label{fig:sparse_weights_distribution}
    \end{subfigure}\\
    \begin{subfigure}{\textwidth}
    	\centering
    	\includegraphics[width=0.8\textwidth]{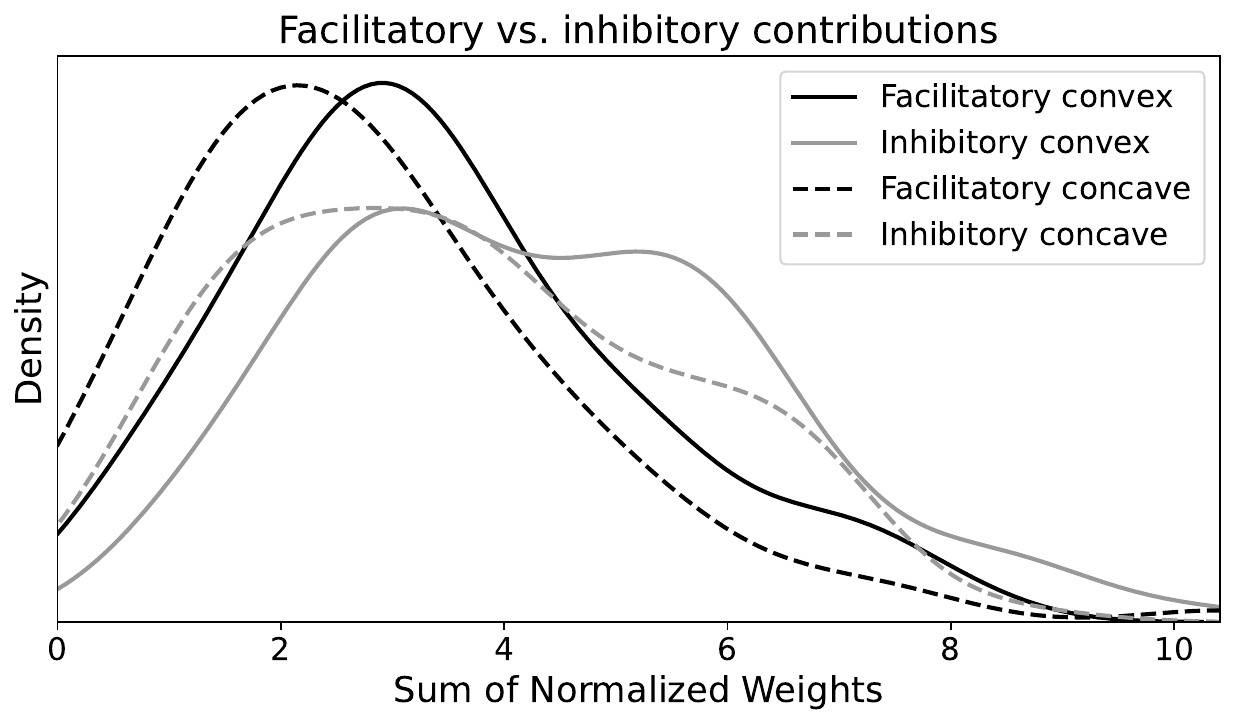}
    	\caption{}
    	\label{fig:exc_inh_weights_distribution}
    \end{subfigure}
 	\caption{Sparsity and contribution type analysis. (\subref{fig:sparse_percentage}) For each neuron in the mV4 population, we calculated the percentage of nonzero convex and concave elements of the sparse code vector. Each neuron is shown with a dot. Dot sizes are proportional to the number of neurons matching the same sparsity. This figure demonstrates that more than a single or three curvature components, such as those modeled  in APC--2D and APC--4D, contribute to V4 activations. Moreover, more dots above the identity line in this plot confirm previous observations of bias toward convexities in V4 neurons~\cite{Pasupathy99_contour_features,carlson2011sparse}. (\subref{fig:sparse_weights_distribution}) Distributions of normalized weights from mLocalCurv maps to mV4 neurons separated according to the eight modeled signed curvatures. Overall, all distributions overlap except for the acute convex weights that peaks at one. This plot in conjunction with results from panel (\subref{fig:sparse_percentage}) indicate a bias toward convexities in mV4 neurons. (\subref{fig:exc_inh_weights_distribution}) Facilitatory versus inhibitory weight distributions. The distribution of facilitatory and inhibitory contributions from shape parts greatly overlap, with a slight shift for facilitatory convex weights, suggesting both types of contributions determine V4 responses.}
 	\label{fig:distributions}
 \end{figure}

To measure the number of parts contributing to mV4 responses in SparseShape, we examined the sparsity of the learned sparse code vector as the percentage of its nonzero elements separately for convex and concave parts for each mV4 cell. Figure~\ref{fig:distributions}(\subref{fig:sparse_percentage}) demonstrates the sparsity distribution for the population with each neuron displayed as a dot. 
This  distribution suggests contributions from more than a single or even three shape parts to mV4 responses. Additionally, convex sparsity median at 0.53 compared to 0.5 for concave sparsity implies a bias toward convexities in agreement with the observations reported in~\cite{Pasupathy99_contour_features,carlson2011sparse}.
 
 With more convex parts contributing to mV4 responses, one possibility is smaller convex weights compared to fewer but larger concave weights to compensate for this imbalance. To evaluate this possibility, for each neuron, we normalized the sparse code vector with maximum set at 1. Then, we took the largest weight from each signed curvature type (each mLocalCurv map), resulting in an eight-element vector for each neuron. Considering the population distribution over each element of this vector, a shift  to larger weights is expected from signed curvature components with substantial contributions to mV4 responses. Interestingly, as illustrated in Figure~\ref{fig:distributions}(\subref{fig:sparse_weights_distribution}), all distributions are relatively overlapping, except for that of acute convexities peaking at one, 
 confirming a bias toward convexities in the learned model. 
 \subsection{Contribution types}
If V4 neurons indeed integrate V2 responses in both excitatory and inhibitory manners, models limited to facilitatory weights neglect accounting for important contributions to V4 responses. 
 In SparseShape, both types of contributions are explored during learning and accounted for in the model. To assess the effect of contribution type to mV4 responses, we considered the population distribution of sum of facilitatory and inhibitory weights in the normalized sparse code vector (max at 1). Interestingly, as depicted in Figure~\ref{fig:distributions}(\subref{fig:exc_inh_weights_distribution}), the distributions plotted separately for convex and concave parts demonstrate larger inhibitory effects from both convex and concave parts, suggesting that excluding inhibitory contributions result in an incomplete understanding of V4 shape processing. To further investigate the effect of inhibitory contributions, we performed an ablation study in which we restricted part contributions to excitatory ones in SparseShape. With this imposed constraint, as demonstrated in Figure~\ref{fig:error_correlation_comparison}(\subref{fig:correlation_comparison}), SparseShape--ablation--train and test correlation distribution medians are decreased to 0.62 and 0.52 respectively. These results, compared to train/test medians at 0.71 and 0.6 in SparseShape with inhibitory part contributions, highlight the importance of inhibition in V4 responses.

 \subsection{RF visualization}
\begin{figure}[!tp]
 	\centering
 	\begin{subfigure}{\textwidth}
 		\centering
 		\includegraphics[width=\textwidth]{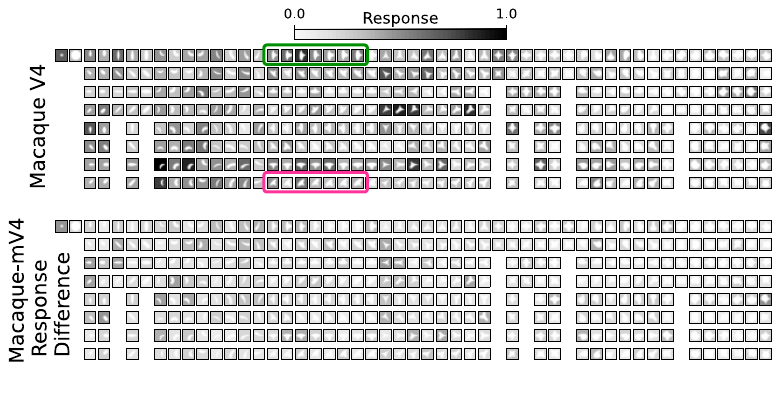}
 		\caption{}
 		\label{subfig:V4519_resp}
 	\end{subfigure}\\
 	\begin{subfigure}[c]{0.5\textwidth}
 		\centering
 		\includegraphics[width=\textwidth]{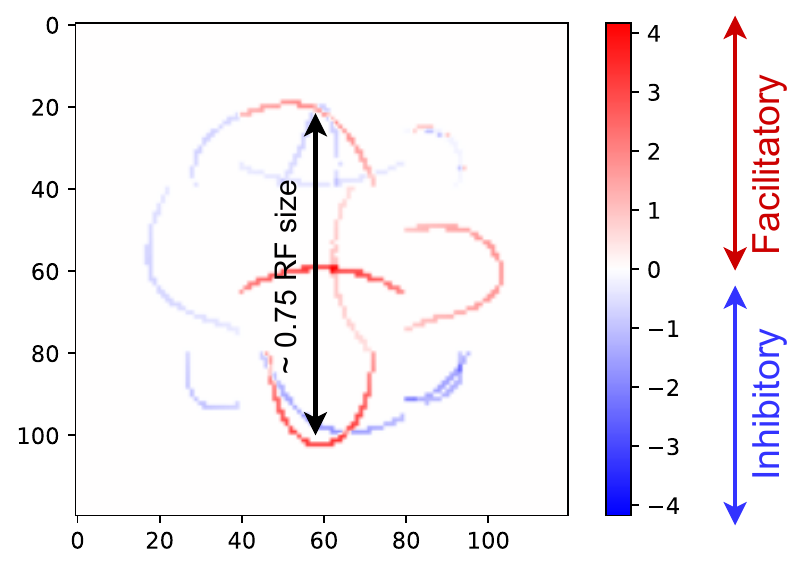}
 		\caption{}
 		\label{subfig:V4519_RF}
 	\end{subfigure}~~
    \begin{subfigure}{0.4\textwidth}
    	\centering
    	\includegraphics[width=\textwidth]{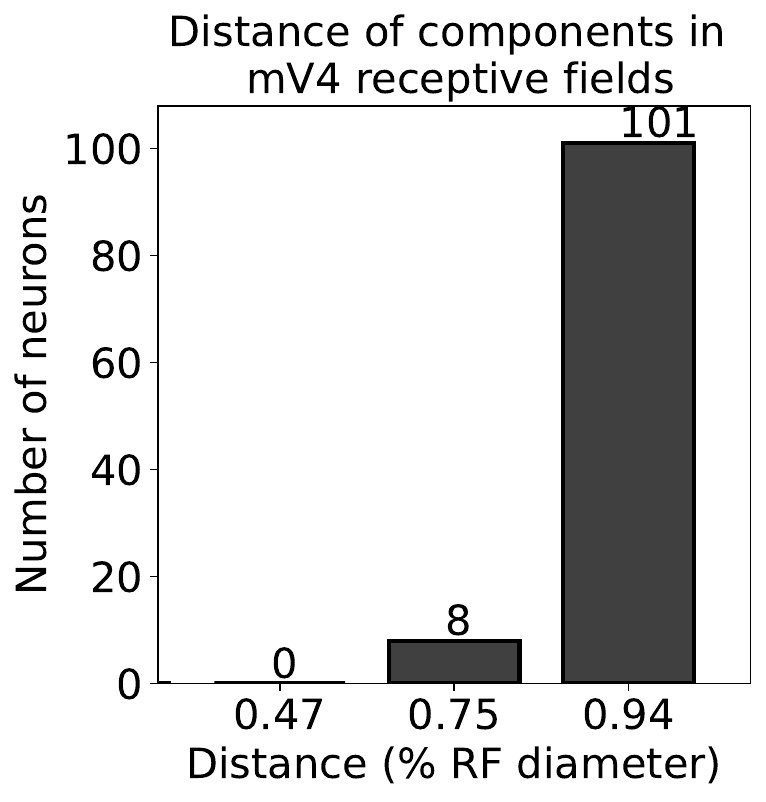}
    	\caption{}
    	 \label{fig:RF_part_distance}
    \end{subfigure}
 	\caption{Recovered mV4 RFs.~(\subref{subfig:V4519_resp}) Responses of a mV4 cell are shown. The shapes inside the green box evoke strong activations in the biological neuron. Same shapes but rotated at -45deg inside the magenta box evoke much weaker activations in this cell.~(\subref{subfig:V4519_RF}) The recovered RF of the same neuron whose responses are shown in panel (\subref{subfig:V4519_resp}). Red and blue indicate facilitatory and inhibitory contributions. Shapes with medium/broad convexities at the top, bottom and right part of the RF evoke strong activations in the cell. In contrast, medium/broad convexities in the bottom right and left parts of the RF inhibit neuronal responses. The recovered RF displays that shape parts from the spatial extent of the neuron's RF, as far as 0.75$\times$ RF size,  are combined to account for its selectivity. Such long-range interactions between shape parts were not captured in the APC model due to fitting a single Gaussian to neuron responses. (\subref{fig:RF_part_distance}) Population histogram of shape part distances within mV4 RFs. The majority of mV4 cells integrate shape parts that are apart at least as far as three forth of the RF diameter, suggesting V4 neurons incorporate V2 responses from the full span of their RF. Note that the .95 $\times$ RF diameter is for shape parts on the diagonal of the simulated square-shaped RFs.}
 	\label{fig:V4519_resp_corr_RF}
 \end{figure}

 Visualizing the recovered receptive fields reveals shape selectivities in each mV4 cell. In SparseShape, modeling a hierarchy of representations that include intermediate and higher-level feature of signed curvature makes explaining the recovered receptive fields effortless. For instance, Figure~\ref{fig:V4519_resp_corr_RF} shows an example neuron responses along with a visualization of its recovered receptive field. 
Figure~\ref{fig:V4519_resp_corr_RF}(\subref{subfig:V4519_RF}) demonstrates selectivity to mild and broad convexities in the top/bottom/right part of the RF. Responses of this cell are inhibited with appearance of mild convexities on bottom-right and bottom-left parts of the RF respectively. 
 The recovered RF can be qualitatively verified with the Macaque V4 responses in Figure~\ref{fig:V4519_resp_corr_RF}(\subref{subfig:V4519_resp}). For example, the shapes within the green rectangle in Figure~\ref{fig:V4519_resp_corr_RF}(\subref{subfig:V4519_resp}) have three convex parts on facilitatory positions causing relatively strong activations of the cell. The same shapes rotated with two convex part within the inhibitory parts of the RF (encompassed with a magenta rectangle) invoke weaker activations in the neuron. 
 
 Figure~\ref{fig:V4519_resp_corr_RF}(\subref{subfig:V4519_RF}) presents an interesting observation: the contributions to this neuron's responses are from the full spatial extent of the cell with interacting shape parts separated at almost 0.75$\times$ RF size. To quantify the extent of shape part interactions within the RF for each neuron, we computed the distance of all pairs of contributing parts in the RF and took the maximum distance among all parts. Figure~\ref{fig:V4519_resp_corr_RF}(\subref{fig:RF_part_distance}) shows the mV4 population distance histogram indicating that the majority of neurons integrate shape parts from at least as far as three-forth of their RF diameter. This observation suggests that a single-Gaussian prior is far too limiting to capture all the factors contributing to V4 responses.

 \subsection{Invariance}
 \label{section:invariance}
\begin{figure}[!tp]
 	\centering
    \begin{subfigure}{\textwidth}
    	\centering
    	\includegraphics[width=\textwidth]{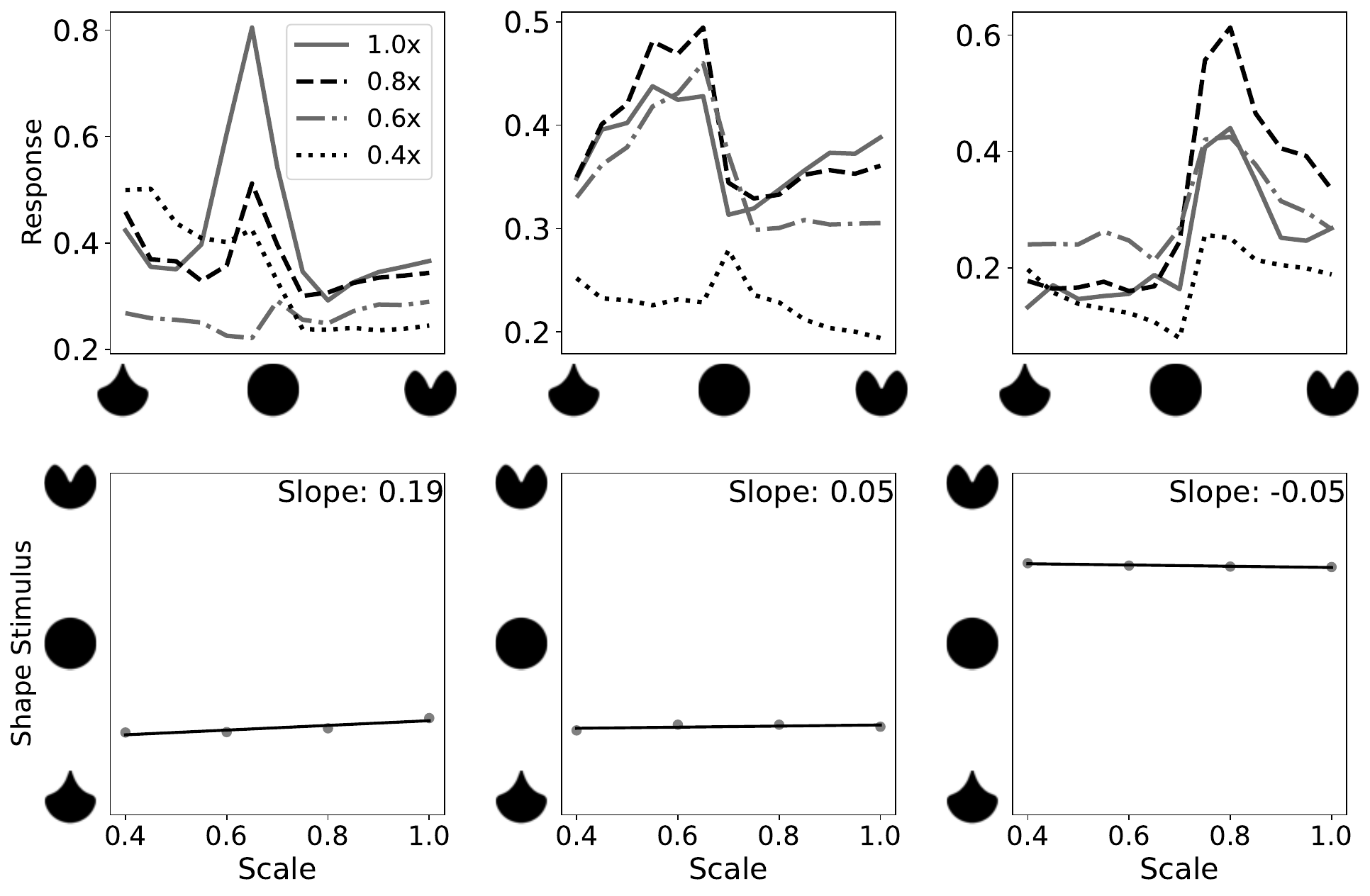}
    	\caption{}
        \label{subfig:scale_inv_sample_resp}
    \end{subfigure}
	 \begin{subfigure}{0.49\textwidth}
	 	\centering
	 	\includegraphics[width=\textwidth]{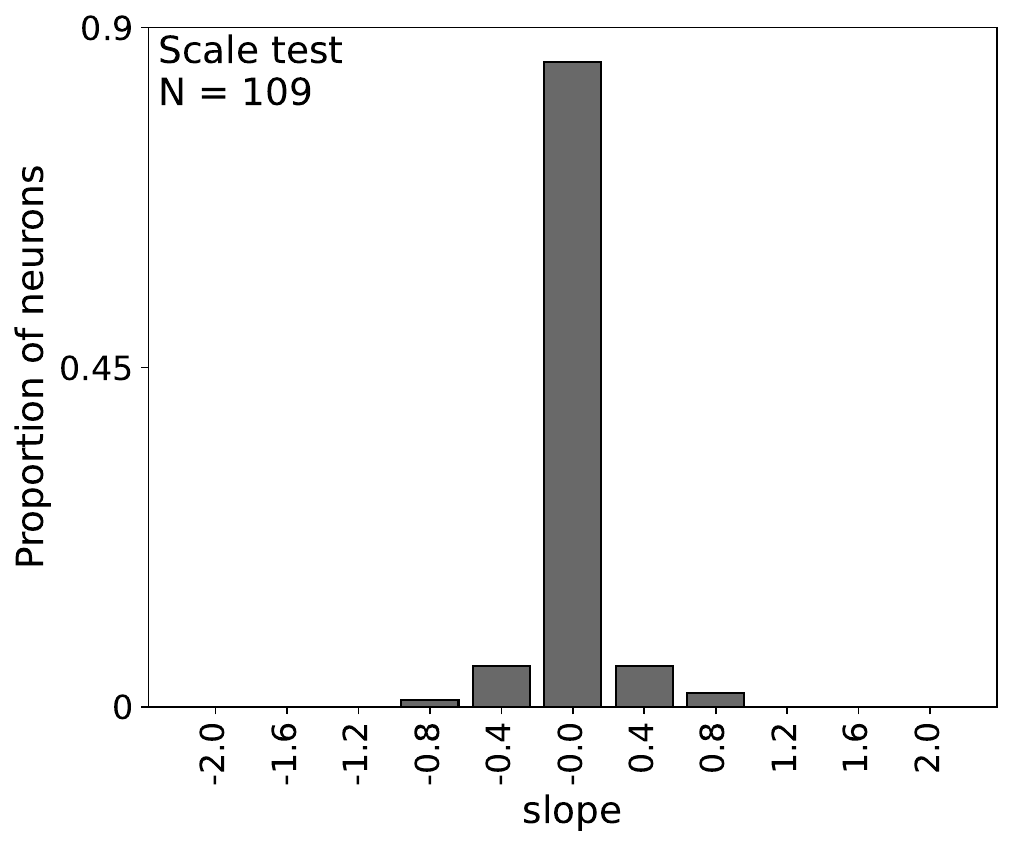}
	 	\caption{}
	 	\label{subfig:scale_model_histogram}
	 \end{subfigure}~
	 \begin{subfigure}{0.49\textwidth}
	 	\centering
	 	\includegraphics[width=\textwidth]{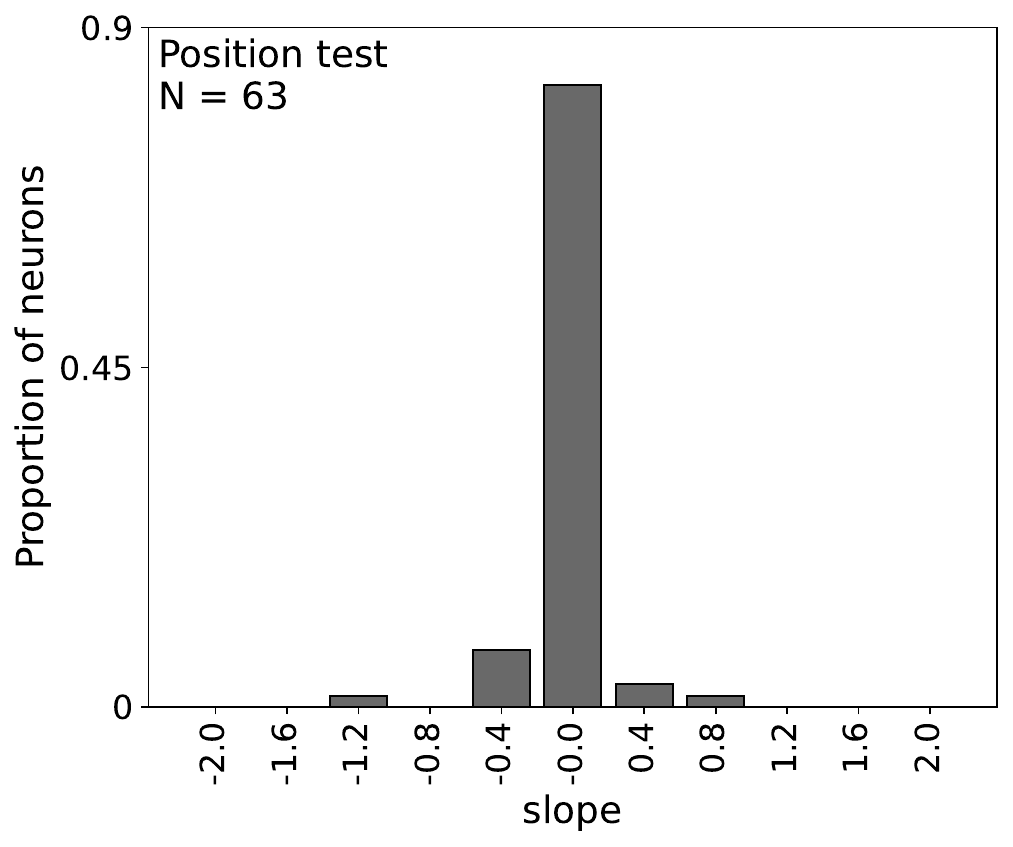}
	 	\caption{}
	 	\label{subfig:scale_model_position_histogram}
	 \end{subfigure}\\
 	\caption{The learned mV4 neurons exhibit scale and position invariance. (\subref{subfig:scale_inv_sample_resp}) Scale-invariance in responses of mV4 neurons. The top row shows responses of three example mV4 cells to stimuli of various scales with their corresponding tuning centroid slope as a function of scale shown in the bottom row in this panel. These plot suggest that mV4 cells exhibit ``normalized'' curvature encoding similar to Macaque V4 neurons. Same legend applies to all plots on the first row. (\subref{subfig:scale_model_histogram}) Tuning centroid slope distribution for the population of mV4 neurons in SparseShape for the scale test.~(\subref{subfig:scale_model_position_histogram}) Slope distribution for the position test. Panels (\subref{subfig:scale_model_histogram}), (\subref{subfig:scale_model_position_histogram}) demonstrate that the learned mV4 neurons show similar scale and position invariance properties as Macaque V4 neurons~\cite{elShamayleh2016scale} despite training on a single scale.}
 	\label{fig:scale_inv_results}  
 \end{figure}

 El-Shamayleh~\etal~\cite{elShamayleh2016scale} reported a normalized curvature encoding in V4 cells. We probed our mV4 cells to determine if the learned selectivities encode normalized or absolute curvature (See~\cite{elShamayleh2016scale} for definitions and details of experimental setting). For this purpose, we prepared shapes of varying scales similar to those employed by El-Shamayleh~\etal~\cite{elShamayleh2016scale} 
 and evaluated systematic shifts in tuning centroids in mV4 responses. Briefly, if these neurons exhibit scale invariance, their tuning peak would not shift with changes in scale. Therefore, measuring the shift in tuning centroids as a function of scale would result in slopes close to zero. To examine if the observed invariance in responses could be attributed to the shift in the position of boundary conformation, we tested our mV4 responses to changes in stimuli position within RF, similar to~\cite{elShamayleh2016scale}. Figure~\ref{fig:scale_inv_results}(\subref{subfig:scale_inv_sample_resp}) demonstrates a few examples of mV4 neuron responses with scale-invariant selectivities as indicated by the small slopes. 
Population histograms of tuning centroid slopes, depicted in Figure~\ref{fig:scale_inv_results}(\subref{subfig:scale_model_histogram}), (\subref{subfig:scale_model_position_histogram}), suggest that the learned mV4 neurons exhibit invariance properties similar to Macaque V4 cells (compare with histograms in~\cite{elShamayleh2016scale}, figures 5 and 9).
 
 Results from this experiment are interesting in that our mV4 neurons were trained with a single scale stimuli set. However, unlike HMAX whose responses undergo substantial changes with variations in scale and position~\cite{bair2015modeling}, the learned weights in SparseShape generalized to account for these variations in responses of Macaque V4 cells.

 \section{Discussion}
 Our goal was to understand the shape signal transformation in the ventral stream up to V4. The present work employed findings of the various visual areas involved in shape processing in this stream along with geometry and machine learning to propose a step-by-step explanation for this transformation. We refrained from employing end-to-end deep neural networks because explaining what is learned in each layer of these deep architectures is still an open problem. 
 Instead, we took an alternative approach and proposed a mechanistic model with explicit algorithmic steps that could explain the role each component plays in the shape signal transformation from orientation selectivity to abstract representations of signed curvature and part-based selectivity in V4. We stood on the shoulder of experimental findings of the brain where possible and designed the model and set its parameters accordingly. Only in absence of such knowledge, we gained from a machine learning algorithm to unlock new information hidden in existing V4 data: our results indicate contributions from multiple parts from the full spatial extent of RF in both facilitatory and inhibitory manners. 
 
 
 Throughout this manuscript, we made the effort to demonstrate the important role both curvature magnitude and sign representations play in achieving a signed curvature encoding. Recently, Pasupathy~\etal~\cite{pasupathy2020visual} referred to V4 representations as ``object-centered''. We, instead, used the geometric term ``signed curvature'' to emphasize selectivity to both curvature magnitude and curvature sign in V4 neurons as was initially reported in~\cite{Pasupathy99_contour_features,pasupathy2001_shape_representation,pasupathy2002_nature}. Object-centered V4 selectivities refers to awareness to inside-outside (border ownership) and omits the curvature magnitude selectivity in these neurons.  
  
  Similarly, when it comes to a model of V4, those previous models that omit incorporating both curvature components cannot achieve a signed curvature encoding and hence, are lacking with respect to our current understanding of shape processing in V4. Even with impressive population distribution of correlation coefficients, HMAX falls under this category. Without incorporating curvature sign, it is not surprising that HMAX failed to exhibit similar levels of invariance to changes in position as V4 neurons~\cite{bair2015modeling}; with an example, we showed (Figure~\ref{fig:signed_curvature}(\subref{fig:HMAX_FG_example})) that the same configuration of oriented edges recovered in HMAX could be a convex or concave segment on the bounding contour of a shape depending on the shape position in the receptive field. Additionally, the HMAX hierarchy skips intermediate processing stages and jumps from orientation-selective neurons to shape selectivity in V4. As a result, some of the recovered RFs look like an ensemble of oriented edges that are difficult to interpret. Such an approach fails to explain how the shape signal transforms in the ventral stream. In contrast, with explicit modeling of intermediate representations, not only did we achieve a signed curvature encoding, but also provided a step-by-step explanation for the development of this representation in the ventral stream.
  
Compared to other models, SparseShape outperformed 2DSIL, Alexnet and APC. Even though APC was fit to all the shapes in the standard set and we trained our model with only 60\% of shapes, SparseShape--train outperformed APC--4D. This mismatch on performance suggests more than three juxtaposed shape parts contribute to V4 responses. This observation was confirmed in our sparsity experiment and was evident in the visualized receptive fields: V4 neurons integrate shape parts from the full spatial extent of their RF, which is more efficient than limiting selectivity to a small portion of the RF. 

Not only long-range interactions of shape parts within the RF emerged from relaxing the priors compared to APC, a combination of facilitatory and inhibitory contributions also appeared in the recovered RFs. All previous models of V4 were limited to facilitatory contributions of parts. However, the relaxed priors in SparseShape revealed more complicated part-based selectivities in V4 neurons. Facilitatory and inhibitory part contributions in SparseShape reiterates the findings of Brincat~\etal~\cite{Brincat2004} in IT and therefore, it is not surprising to find that V4 neurons follow suit.
 
In SparseShape, the signed curvature encoding relies on both endstopping and border ownership. Discoveries of both types of representations in the ventral stream lend additional support to the development of a signed curvature encoding in this stream. Responses of V4 cells to shapes that are extended beyond the RF~\cite{Pasupathy99_contour_features} and similar responses in border ownership cells suggest similar mechanisms govern activations in both neuron types. We acknowledge that findings of Bushnell~\etal~\cite{bushnell2011partial} suggest against border ownership contributing to V4 responses. Their argument is on the basis of time course in BO and V4 neurons. SparseShape incorporates the RBO network to generate the BO signal. The RBO hierarchy is designed based on early recurrence from MT to border ownership neurons and provides an explanation for the early divergence of responses in BO cells~\cite{RBO}. Whether this early divergence is relayed to V4 cells, earlier than reaching half-max responses in BO as used for the argument in~\cite{bushnell2011partial}, or a direct recurrence from MT provides side-of-figure information to V4 neurons remains to be further investigated. The message here, however, is that both curvature magnitude and curvature sign (perhaps from similar mechanisms that give rise to border ownership) are essential to forming a signed curvature encoding. How and where from in the brain this information is provided to V4 is another question that is put forth to be investigated in the future. 

Our proposed model revealed hidden information, more than those previously reported, in existing V4 data. These findings impart new insights into shape processing in V4 that require further testing and investigation in biological V4 neurons. For example, examining biological V4 responses by removing shape parts allow testing for facilitatory versus inhibitory contributions, multi-part selectivity and the extent of part integration within the receptive field. Removing a shape part inevitably disturbs the curvature sign in a geometric sense and might make the experimental design a challenge in this case. Luckily, Zhang~\etal~\cite{BOwn_Zhang2010} found that the side-of-figure preferences are maintained in border ownership neurons even when boundary fragments are removed from Cornsweet shapes. Additionally, the RBO network could explain illusory contours. Together, these findings provide a unique tool to further test the signed curvature encoding in the ventral stream and to enhance our understanding of contributions of shape parts to V4 responses. Specifically, if V4 cells receive inside-outside information that originates from BO cells, converting the standard stimulus set to Cornsweet shapes allows removal of shape parts without disturbing the overall inside-outside and consequently the signed curvature signal for the shape. Then, the findings from the present work can be examined in biological V4 cells. 
For example, removing a part that inhibits a cell's responses according to the model is expected to increase the neuron's activations when all other shape parts are intact. Note that such a study is different from those in which V4 responses to occlusion were examined~\cite{bushnell2011partial,fyall2017occlusion}. In the occlusion studies, the shape is occluded with another form which disturbs border ownership and consequently signed curvature representations. In contrast, with the suggested experimental design, there is no occlusion but absence of a part, making testing the effect of each individual part a possibility without changing the overall shape representation.

Our proposed model can be further extended. For example, in the present model, we combined responses of border- and edge-selective mBO cells. Maintaining those signals allows modeling a wider variety of mLocalCurv neurons and consequently modeling V4 responses to outline/filled shapes~\cite{popovkina2019outline}. Similarly, adding neurons selective to texture could lend insight into joint shape and texture processing in this areas~\cite{V4_Kim_shape_texture}. Finally, we did not implement mechanisms to represent inflection points (zero curvature) such as straight lines at mV4 level. Adding connections from mV1 layer to mV4 can handle such cases. 
Such extensions are left for future work. 

\section{Online Methods}
\label{section:Methods}
The SparseShape network combines and extends 2DSIL~\cite{Antonio_2012} and the RBO network~\cite{RBO}. Whereas 2DSIL models curvature magnitude and direction (the direction toward which the contour curves) through endstopped neurons, the RBO network provides figure-ground information. Combining these features yields a unique curvature sign for each point on the bounding contour of a shape within the visual field. 

The SparseShape architecture depicted in Figure~\ref{fig:model_architecture}(\subref{subfig:hierarchy}) represents each model neuron type in the hierarchy with a single box. Similar to 2DSIL and RBO, simple and complex cells are selective to oriented edges. Model border ownership (mBO) neurons signal figure side. Curvature magnitude and direction are encoded at the endstopped level in model endstopped degree (mEsDeg) and model Endstopped direction (mEsDir) neurons. Curvature sign (mEsSign) is new in SparseShape and combines mBO and mEsDir signals. The local curvature (mLocalCurv) neurons from the original 2DSIL are remodeled to receive input from mEsDeg and mEsSign cells and encode \textit{signed curvature} by definition. Finally, the arrow in the last network layer represents the set of weights that are learned by employing a supervised sparse coding algorithm. 

SparseShape implements edge- and border-selective neurons at 12 orientations in $[0, \pi)$ and combines responses of mBO neurons with edge- and border-selectivity at the same orientation to a single inside-outside signal that is fed to mEsSign cells. Following 2DSIL and RBO, SparseShape implements neurons at four scales that result in eight mLocalCurv maps (four scales representing curvature magnitude $\times$ two signs). Additionally, neurons up to and including mBO, mEsDeg and mEsDir are the same as those in RBO and 2DSIL, which computational details can be found in~\cite{RBO, Antonio_2012}. Below outlines the computations of new and re-modeled neurons in SparseShape and the supervised learning step.

 \subsubsection{Model Endstopped - Curvature Sign}
 Intuitively, a contour segment of a simple closed curve is convex when the contour segment curves toward inside the shape. Figure~\ref{fig:signed_curvature}(\subref{subfig:curv_sign_two_arrows}) gives examples for which curvature direction  and inside-outside information determine curvature sign for contour segments as parts of two shapes. Both curvature direction and inside-outside information have geometric interpretations (See~\cite{pressley2010elementary} for details): for each point on a contour segment, curvature direction denotes the direction of the tangent vector derivative whereas inside-outside information represents the direction of the unit normal to the curve. When the tangent derivative and unit normal have similar directions, the signed curvature is positive and the curve is convex; otherwise, the contour segment is concave (assuming exclusion of inflection points). In our proposed hierarchy, curvature direction and inside-outside signal are modeled in mEsDir and mBO cells, together making the curvature sign modeling possible in this network.
 
In SparseShape, at each visual field location, a pair of mBO neurons with identical local feature selectivity but opposite side-of-figure preferences are modeled. Similarly, a pair of mEsDir with opposite direction selectivities but identical orientation are modeled. Between each pair, the neuron with stronger response is called the winning cell, for example, the winning mBO neuron. When the direction of the winning mEsDir and mBO neurons are in agreement, the contour segment is convex and concave otherwise, resulting in the following implementation of mEsSign cells:
 \begin{eqnarray}
 R_\text{mEsSign\_convex}(x,y, \theta) &=& \phi((R_\text{mEsDir}(x, y, \theta + \frac{\pi}{2}) - R_\text{mEsDir}(x, y, \theta - \frac{\pi}{2})) \times \nonumber\\
 &&(R_\text{mBO}(x, y, \theta + \frac{\pi}{2}) - R_\text{mBO}(x, y, \theta - \frac{\pi}{2}))), \nonumber\\
 R_\text{mEsSign\_concave}(x, y, \theta) &=& \phi(-(R_\text{mEsDir}(x, y, \theta + \frac{\pi}{2}) - R_\text{mEsDir}(x, y, \theta - \frac{\pi}{2})) \times \nonumber\\
 &&(R_\text{mBO}(x, y, \theta + \frac{\pi}{2}) - R_\text{mBO}(x, y, \theta - \frac{\pi}{2}))),
 \end{eqnarray}
 where $(R_\text{mEsDir}(x, y, \theta + \frac{\pi}{2}) - R_\text{mEsDir}(x, y, \theta - \frac{\pi}{2}))$ and $(R_\text{mBO}(x, y, \theta + \frac{\pi}{2}) - R_\text{mBO}(x, y, \theta - \frac{\pi}{2}))$ denote the direction of winning mEsDir and mBO neurons and the multiplication tests their direction alignment. $R_{x}$ represents the response of neuron type $x$ with $x\in\{\text{mEsSign\_convex},\text{mEsSign\_concave},\\ \text{mEsDir},\text{mBO}\}$ and $\theta$ denotes the orientation selectivity of the corresponding mEsDir or mBO cell. The function $\phi$ rectifies negative responses to zero.
 
\subsubsection{Model Local Curvature}
Similar to mEsDir and mEsSign cells, \textit{signed curvature} is encoded by a pair of mLocalCurv neurons at each visual field location to represent positive and negative signs for a given curvature magnitude. Specifically, at each scale in the hierarchy, a single map of mEsDeg and a pair of mEsSign (mEsSign\_convex and mEsSign\_concave) are combined to yield pairs of mLocalCurv cells as:
 \begin{eqnarray}
 R_\text{mLocalCurv\_pos}(x, y) & = & \max_{j=1}^{12}(R_\text{mEsDeg}(x, y, \theta_j) \cap \nonumber\\
 &&(R_\text{mEsSign\_convex}(x,y, \theta_j) > R_\text{mEsSign\_concave}(x,y, \theta_j))),\nonumber\\
 R_\text{mLocalCurv\_neg}(x, y) & = & \max_{j=1}^{12}(R_\text{mEsDeg}(x, y, \theta_j) \cap \nonumber\\
 &&(R_\text{mEsSign\_convex}(x,y, \theta_j) < R_\text{mEsSign\_concave}(x,y, \theta_j))).
 \end{eqnarray}
where ``$\max$'' retains responses to curvature at this level of the hierarchy and marginalizes orientation. Consequently, the final layer feeding to mV4 neurons consists of eight curvature maps, four scales $\times$ two signs, representing four levels of acuteness for convexities and concavities.
 
 \subsection{Model V4: Learning Receptive Fields}
Our goal was to learn V4 receptive fields such that complex and long-range interactions between shape parts, if they exist, can be captured. Recovering existence or equivalently lack of such interactions in V4 imparts significant insight into shape processing mechanisms in this visual area. 

We trained the weights in the last layer of SparseShape, mLocalCurv to mV4, and recovered the receptive fields from the recordings provided to us by Dr. Anitha Pasupathy. Briefly, for each Macaque V4 cell, we assigned its responses to a mV4 cell in SparseShape and learned the weights from mLocalCurv cells to each mV4 neuron. That is, the procedure for learning the RF was repeated for each individual mV4 cell. 

A naively-added fully-connected layer from mLocalCurv to mV4 cells has more than 14K weights to learn from less than 366 data points. To compensate the imbalance between the number of parameters and data, we propose imposing sparsity priors that are compatible with discoveries of V4~\cite{carlson2011sparse} and other brain areas involved in shape representation~\cite{tsunoda2001complex}. 
We leverage sparsity in a higher dimensional space and with a more relaxed model compared to APC. 
With the learned RFs, obtaining mV4 responses to any arbitrary stimuli set, such as shapes in the invariance experiment explained in Section~\ref{section:invariance}, is a simple feedforward pass (with dorsal recurrence) in SparseShape.

Our proposed sparse coding method formulates RF recovery as a supervised learning problem. Specifically, given the responses of mLocalCurv cells and Macaque V4 responses to a stimuli set, we seek a sparse combination of curvature components across the RF that can explain observed responses by mininizing the following objective function:
\begin{equation}
	\min_{\gamma} \sum_{t = 1}^\tau \frac{1}{2}\|R_t - D_t^T \gamma\|_2 + \alpha \rho \|\gamma\|_1 + \frac{\alpha (1 - \rho)}{2}\|\gamma\|_2^2,
	\label{eq:sparse_optimization}
\end{equation}
where $R_t$ is the Macaque V4 response to stimulus $t$ and $\tau$ is the number of stimuli in the training set. This objective function, known as the elastic net model, combines L1 and L2 penalties to ensure a sparse representation with regularized learned weights. In this equation, $D_t$ is the part-based vector whose elements signal presence/absence of a shape part at a particular position within the RF and $\gamma$ is the sparse code vector that specifies the weight of each curvature component contributing to responses. The sign of each element in $\gamma$ determines facilitatory versus inhibitory contribution. The trade-off between L1/L2 norms and the error term is enforced through $\alpha$. $\rho$ determines the balance between L1 and L2 norms. Both $\alpha$ and $\rho$ are set with cross-validation. 

The part-based vector $D_t$ is obtained by a set of Gaussian kernels that filter mLocalCurv maps. In particular, to account for a variety of shape part positions, a 3$\times$3 grid over the RF filters each mLocalCurv map. Each cell of this grid encompasses a Gaussian kernel. The parameters of each Gaussian characterize the position and extent of a particular curvature component within a cell. Putting these together, $d_t(i)$, the $i$-th element of $D_t$, is computed as: 
\begin{equation}
	d_t(i) = \sum_{x, y} \mathbf{C}(x, y,t, k) \cdot \mathbf{G}(x,y,j; \mu_j, \Sigma_j),
	\label{eq:dictionary}
\end{equation}
with $\mathbf{C}(x, y, t, k)$ corresponding to the responses in the $k$-th mLocalCurv map to stimulus $t$ at visual location $(x,y)$ with $k\in\{1,2,\ldots,8\}$. In this equation, $\mathbf{G}(x, y,j; \mu_j, \Sigma_j)$ represents the Gaussian in the $j$-th grid cell with parameters $\mu_j,\Sigma_j$ and $j\in\{1, 2, \ldots, 9\}$. We explored learning the parameters of the Gaussian kernels for each neuron. However, we did not find any benefits in learning these kernels compared to fixing those to pre-determined kernels. As such, we fixed the parameters of the Gaussians in this work. With 8 mLocalCurv maps and 9 cell positions, $D_t$ and $\gamma$ in our implementation are 72-dimensional vectors. 

Having found the desired curvature components and their weights, responses of the mV4 neurons corresponding to a given Macaque V4 cell to an arbitrary stimulus $s$ can be obtained by:
\begin{equation}
	R_\text{mV4} = D_s^T\gamma.
	\label{eq:V4_resp_equation}
\end{equation} 

The diagram in Figure~\ref{fig:model_architecture}(\subref{subfig:sparse_model}) shows the different components of our sparse coding formulation. The green box in this figure corresponds to the green arrow in Figure~\ref{fig:model_architecture}(\subref{subfig:hierarchy}).

\bibliographystyle{ieeetr}
\bibliography{./bibliography, ./longstrings}

\end{document}